\documentclass[prl,twocolumn,showpacs,superscriptaddress]{revtex4}
\usepackage{graphicx}
\usepackage{dcolumn}% Align table columns on decimal point
\usepackage{bm}% bold math

\begin{document}
\title{Dynamics of metallic stripes in cuprates}
\author{J. Lorenzana}
\affiliation{CONICET, Centro At\'omico Bariloche,
8400 S. C. de Bariloche, Argentina and INFM Center for Statistical 
Mechanics and Complexity,  
Universit\`a di Roma ``La Sapienza'', P. Aldo Moro 2, 00185 Roma, Italy}
\author{G. Seibold}
\affiliation{Institut f\"ur Physik, BTU Cottbus, PBox 101344,
03013 Cottbus, Germany}
%\maketitle
\date{\today}
\begin{abstract}
We study the dynamics of metallic vertical stripes in cuprates
within the three-band Hubbard model based on a recently developed
time-dependent Gutzwiller approximation. As doping increases the
optical conductivity shows transfer of spectral weight from the
charge transfer band towards {\it i}) an incoherent band centered at 1.3eV,
{\it ii}) a Drude peak, mainly due to motion along the stripe,
{\it iii}) a low-energy collective mode which softens with doping and
merges with {\it ii}) at optimum doping in good agreement
with experiment.  The softening is related to the
quasidegeneracy  between Cu centered and O centered mean-field
stripe solutions close to optimal doping.
\end{abstract}
\maketitle

%\begin{multicols}{2}
The doping-dependent evolution from insulating behavior to a strange
metal in the superconducting  cuprates emerges
dramatically in the normal state optical conductivity.
Slightly doped cuprates  show a small or no Drude peak and
doping-induced transfer of spectral weight from the
charge-transfer ($\sim $2eV) to a mid-IR (MIR) band at
$\sim $0.5eV\cite{uch91,suz89}. Upon further doping the system
progressively metallizes as is evident from the prominent 
Drude-like peak that develops at zero energy. A remarkable effect of
doping is that the MIR band strongly softens and merges with the
Drude peak, resulting in a feature that cannot be fitted by a
conventional Drude model. A variety of alternative theories
\cite{alte,cap02} have been proposed in order to describe this
feature. 
Clearly the identification of this low-energy MIR (LEMIR) band is of
paramount importance to understand the physics of these materials.
The low-doping behavior has been explained in terms of the 
random-phase-approximation (RPA)
electronic excitations of single-hole Hartree-Fock states in
CuO$_{2}$ layers\cite{lor93a}, but the moderate doping 
behavior (the softening of the LEMIR band) could not be explained due
to difficulties with the HF ground state.

The softening of the LEMIR band in La$_{2-x}$Sr$_{x}$CuO$_{4}$
(LSCO) is accompanied by the appearance of another (much less discussed)
band at 1.3eV\cite{uch91,suz89,ter99}. This high energy MIR (HEMIR) band is 
  well pronounced in optical absorption through LSCO thin films\cite{suz89},
and electron energy loss spectroscopy\cite{ter99} where it develops as a 
function of doping.  Moreover LEMIR and HEMIR are also
detected in photodoped experiments on LSCO\cite{gin88}.
The HEMIR has not been clearly resolved by reflectivity in 
YBa$_{2}$Cu$_{3}$O$_{6+\delta}$ (YBCO),
but a strong broad feature at the right energy appears
in photodoped transmission experiments\cite{len91}. 
As far as we know no microscopic explanation of the HEMIR exists
so far.

Another important aspect of layered cuprates that has emerged in
the last years is the rearrangement of doped holes in
antiferromagnetic (AF) domain
walls\cite{tra95,tra96,tra97,yam98,ara99,ara00}. These one-dimensional 
(1D) structures called stripes were predicted by
mean-field theories\cite{zaapolmacsch}.

Recently we have presented a computation of metallic mean-field
stripes\cite{lor02} within an unrestricted Gutzwiller
approximation\cite{gut65} (GA).
The behavior of the magnetic incommensurability 
$\epsilon=1/(2d)$\cite{tra97,yam98,ara99,ara00}
($d$ is the distance between charged stripes in units of the
lattice constant), chemical
potential\cite{ino97,har01}, and transport
experiments\cite{nod99,wan01} as a function of doping have been
explained in a parameter free way\cite{lor02}.

\begin{figure}[tbp]
\includegraphics[width=7cm,clip=true]{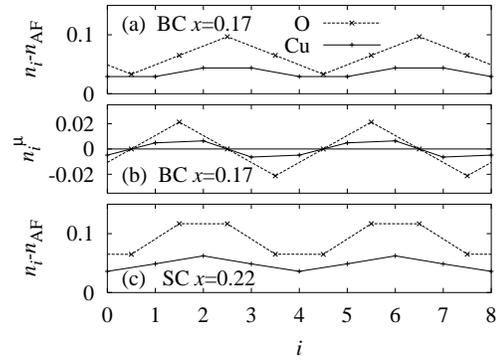}
\caption{Charge density minus undoped charge density as a
 function of atomic position in the direction
  perpendicular to the $d=4$ stripe for the BC solution at $x=0.17$ (a) and
  the SC solution at $x=0.22$ (c). (b) shows the transition charge for the
 LEMIR collective mode of the panel (a) solution.  }
\label{fig:rdx}
\end{figure}
\begin{figure}
\includegraphics[width=7cm,clip=true]{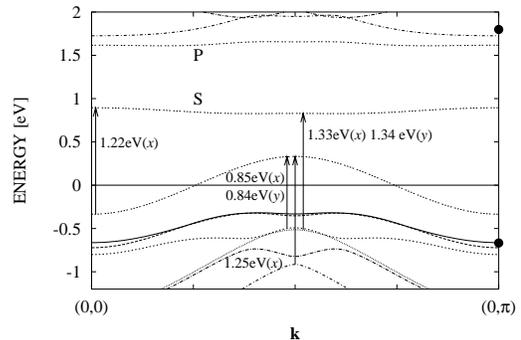}
\caption{Electron mean-field bands measured from the chemical potential
 for the momentum in the direction of the stripe ($d=7$, $x=0.071$).
The arrows are the lowest energy dipole allowed mean-field transitions
in a $d\times4$ cluster labeled by energy and polarization.
Notice that the $x$ polarization is perpendicular to the stripe.
 We also  plot the energies of the AF insulating bands at momentum
${\bf k}=(0,\pi)$ (full dots) measured from the same reference energy }
\label{fig:edk}
\end{figure}
\begin{figure}[tbp]
\includegraphics[width=7cm,clip=true]{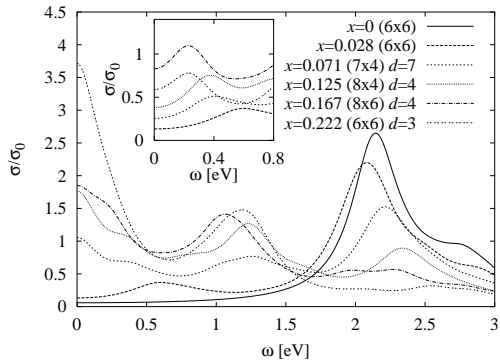}
\caption{Optical conductivity labeled by doping, system size and, in
  the case of stripes, interstripe distance.
 The units of conductivity are given by $\sigma_0=3.6
\times 10^2 (\Omega {\rm cm})^{-1}$ with a background
dielectric constant $\epsilon_b=2.6$ (see Ref.~\protect\cite{lor93a} and
caption of Fig.~\protect\ref{fig:swdx}). The curve labeled
$x=0.028$ corresponds to the single-hole solution. For
larger dopings the figure is an average over the electric field directions
parallel and perpendicular to the BC stripes. The inset shows the low energy
spectra excluding the Drude component. We used a Lorentzian broadening
 of 0.2eV.}
\label{fig:sdw}
\end{figure}
\begin{figure}[tbp]
\includegraphics[width=7cm,clip=true]{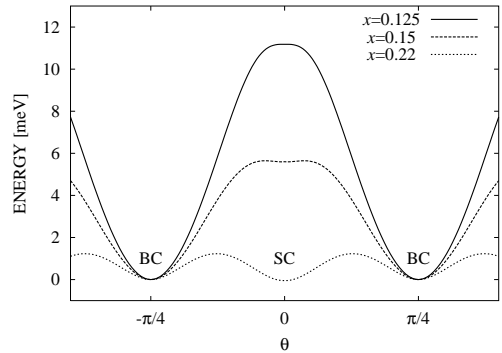}
\caption{Mean field energy per stripe cell (8 Cu's)
as a function of the collective phase $\theta$ of the CDW
for various doping levels and for $d=4$ stripes in a
$4d\times 12$ system.}
\label{fig:edy}
\end{figure}

In this work we investigate the optical conductivity in terms of
linear excitations around the metallic mean-field stripes of
Ref.~\cite{lor02}. RPA
fluctuations are added through a recently developed time-dependent
GA labeled GA+RPA\cite{sei01}.
The stripes show a strong Drude component in the stripe direction,
of magnitude similar to the 
experimental one. In addition the single-hole\cite{lor93a}
 MIR band splits into two bands in good accord with 
the HEMIR and LEMIR (Fig.~\ref{fig:sdw}).
 The  HEMIR band can be understood in terms of incoherent 
transitions between band
states of a stripe at mean-field level (Fig.~\ref{fig:edk}). The LEMIR
band is a collective mode associated with soft
lateral displacements of the stripe.  As doping increases the
collective mode shifts to lower energy and merges with the Drude
part, providing an explanation for the softening of the LEMIR band
in cuprates.
This softening is due to the quasidegeneracy between O-centered
 and Cu-centered stripes found in many 
 approaches\cite{whi98prl80,fle00,lor02} 
and the small energy barrier between these solutions 
(Fig.~\ref{fig:edy}). 
The soft collective mode is a good candidate to mediate pairing between holes
and to produce non-Fermi liquid anomalies in the normal state.
A related scenario was presented in Ref.~\cite{lor93a} but the origin
of the softening was not clear.

We use a three-band model for cuprates with the same LDA 
parameter set\cite{mcm90} as in Ref.~\cite{lor02}. 
%following LDA
%parameter set\cite{mcm90}: $\epsilon_{p}-\epsilon_{d}=3.3$ eV for
%the splitting between the diagonal energies of a hole in the
%copper $d$- and oxygen $p$ orbital, $t_{pd}=1.5$ eV 
%($t_{pp}=0.6$ eV) for the $p$-$d$ ($p$-$p$) hopping
%integral, $U_{d}=9.4$ eV ($U_{p}=4.7$ eV) for the
%repulsion between two holes on the same Cu (O) orbital and
%$U_{pd}=0.8$ eV for the Cu-O repulsion\cite{mcm90}. The largest
%interaction $U_{d}$ is treated within the GA while the others are
%decoupled via the Hartree-Fock (HF) approximation. Our real space
%computations are completely unrestricted except for the spins
%which are assumed to be oriented along the $z$ direction.

We start by reviewing some mean-field
results\cite{yon92,lor93a,lor02} in order to illustrate the
doping dependent charge and spin textures on which the following
computations are based. At very dilute doping ($x \lesssim 0.03$)
due to the long-range Coulomb interaction (not included in our
calculations) each hole will be close to an acceptor preventing
the formation of stripes.  The lowest energy  one-hole solution
consists of a self-trapped state similar to the Zhang-Rice state\cite{zha88}
as found in HF\cite{yon92,lor93a}.

 As doping increases, the donor potential becomes more uniform and 
 screened, favoring the formation of stripes. Experiment shows
that stripes are parallel to the Cu-O bond except 
 at dopings $0.03\lesssim x\lesssim 0.05$ 
where diagonal stripes have been observed\cite{mat00}. 
 Those may be an intermediate state between the isolated polarons 
and the vertical stripes and probably also
require the long-range Coulomb interaction to be stable.
For simplicity we skip this phase and consider vertical metallic solutions.
Weak one-dimensional instabilities probably relevant at low temperatures
are intentionally suppressed in our clusters due to finite size effects.  
    
 In Ref.~\cite{lor02} we have shown that the most
favorable low-doping metallic mean-field  stripe is centered on O
sites bridging two vertical Cu legs, denoted as bond-centered (BC)
in one band models\cite{whi98prl80,fle00}. Fig.~\ref{fig:rdx} (a)
shows a cross section of the charge modulation perpendicular to
the stripe. For larger concentrations the BC solution becomes
degenerate with the site centered (SC) one at $x_0\approx
0.21$\cite{lor02} whose charge profile is shown in
Fig.~\ref{fig:rdx}(c).

The mean-field bands for a BC stripe lattice as reported in
Ref.~\cite{lor02} are shown in Fig.~\ref{fig:edk}. Here we discuss
in more detail the symmetry which determines the selection rules
for optical transitions. Roughly speaking the flat bands labeled S
and P correspond to symmetric (S) and antisymmetric (P)
combinations of orbitals centered on the two legs of Cu that form
the core of the stripe (sites 2 and 3 in Fig.~\ref{fig:rdx}). The
band crossing the chemical potential (hereafter the ``active
band'') is due to the orbitals centered on the core O leg of the
stripe [at 2.5 in Fig.~\ref{fig:rdx}(a)]. The antisymmetric Cu
orbital combination mixes with the core O orbital pushing upwards
(downwards) the P band (active band) close to the edge of the
stoichiometric bands (marked by the full dots).
The lower bands are of mainly O character.
 All optical transitions in the $x$ direction indicated in
Fig.~\ref{fig:edk} are of course between even and odd states with
respect to the stripe central axis.

Dynamical properties are computed within a real space
implementation of the  GA+RPA method\cite{sei01}. In this context
it is essential to use GA+RPA rather than conventional HF+RPA
theory due to the above-mentioned fact that the starting point
mean-field solution matches the experimentally observed
doping-induced incommensurability. As a bonus 
GA+RPA has been shown to be more accurate than HF+RPA\cite{sei01}.
Note that stripes in LSCO show a dynamic character on a scale of a few 
meV\cite{yam98} which is not capture by our starting-point 
mean-field ground state. 
We expect this to affect the spectra at very low energies as 
discussed in Ref.~\cite{cap02} but not on the scale of the 
transitions reported Fig.~\ref{fig:sdw}.

Fig.~\ref{fig:sdw} reports the optical conductivity within GA+RPA
for various dopings.  The RPA optical conductivity for the
single-hole solution appropriate at low doping, as discussed above,
reproduces the results of HF+RPA\cite{lor93a}: Formation of a
doping induced MIR band close to 0.5eV and doping induced transfer
of spectral weight from the charge transfer band to the MIR region
in agreement with experiment in this doping range\cite{uch91}.

For distant stripes ($d=7$) the single-hole MIR band now splits
into two bands. The one at higher energy is a band of incoherent
particle-hole excitations close to 1.3eV which provides a
theoretical explanation for the HEMIR. The position of this band
is nearly independent of doping and can be understood in terms of
transitions within  the stripe band structure at
mean-field level shown in Fig.~\ref{fig:edk}. Indeed the HEMIR is
mainly formed by the 1.33eV($x$) and 1.34eV($y$) mean field
transitions with similar oscillation strengths. This negligible 
renormalization of the mean field transitions by the RPA is
characteristic for incoherent particle-hole excitations. The other
MIR band, shown also in the inset of Fig.~\ref{fig:sdw}, is a
low-energy collective mode and has no mean-field counterpart. In fact 
the 0.85eV($x$)  transition which has a strong oscillator
strength in mean-field does not show up in RPA and instead the 
low-energy collective mode appears which is also polarized
perpendicular to the stripes. A similar mode was found in a study
of stripes within the $t-t'-t''-J$ model \cite{toh99}. The other
transitions reported in Fig.~\ref{fig:edk} have much smaller
spectral weights.

To characterize the LEMIR band  we compute the so called transition charges
$n_i^{\mu}\equiv \langle 0|\hat n_i|\mu \rangle$
where $\mu$ labels RPA excitations.
 $n_i^{\nu}$ is proportional to the
time-dependent charge fluctuation $\delta n_i$ that would occur at
frequency $\omega_{\nu}$ if the state $\nu$ were weakly excited\cite{rin80}.

In  Fig.~\ref{fig:rdx}(b) we show the charge fluctuation
associated with the LEMIR mode. It is
very similar to the difference in charge density between the BC
and SC stripe, indicating that it corresponds to lateral
displacements of the stripe. In fact, if we approximate the charge
modulation by $\cos(q_{\rm cdw} r + \theta)$ with $q_{\rm
cdw}=2\pi(2\epsilon,0)$, this oscillation can be interpreted as a
time-dependent fluctuation of the phase  $\theta$ and thus the
LEMIR excitation can be identified with a phason.
Optically active phasons have zero momentum, but naturally  
a band of phasons exist with a well-defined dispersion
relation.
In continuum models phasons are massless Goldstone modes whereas here the
commensurability of $q_{\rm cdw}$ with the lattice makes them  
have a finite energy at zero momentum.  
This energy however is small and decreases as doping increases. 
 Since the BC and SC state become quasidegenerate at 
$x_0$ it is natural to expect that the phason softening is related to 
the quasidegeneracy  between these states.

In order to substantiate this idea we compute the energy landscape
for ``intermediate'' solutions constrained to be stripes centered
between Cu and O. For each intermediate state we perform a Fourier
analysis and extract the phase $\theta$ of the first Fourier
component of the  charge-density-wave 
(CDW) modulation. Fig.~\ref{fig:edy} shows the
energy for $d=4$ stripes and different dopings. This
provides an upper bound for the energy along the path connecting
BC ($\theta=\pi/4$) and SC ($\theta=0$) solutions where $\theta$
plays the role of a collective coordinate. The curves are periodic
in $\theta$ with period $2\pi/d$ corresponding to a translation by
one elementary unit cell. Remarkably the curve 
acquires an extra periodicity close to optimum doping corresponding 
to the previously found quasidegeneracy between SC and BC 
solutions\cite{whi98prl80,fle00,lor02}.

RPA is essentially an harmonic approximation of the energy
landscape around the mean field solution (SC or BC). The energy 
squared of
each RPA mode is proportional to the curvature (or ``stiffness'') of the
corresponding parabolic energy approximation when the system is
displaced from the stationary state in the direction of the mode
eigenvector (in our case parametrized by the collective coordinate
$\theta$). Fig.~\ref{fig:edy} shows that as doping increases the
stiffness decreases showing explicitly that the softening of the
LEMIR feature is due to the quasidegeneracy  between BC and SC
stripes. Of course the problem is very anharmonic close to $x_0$
and RPA provides only a rough estimate to the phason energy.
Moreover, since the barrier is strongly reduced close to $x_0$ we
expect that anharmonic corrections will make the phason even
softer and the ``true'' ground state will be a fluctuating mixture
of BC and SC solutions.

\begin{figure}[tbp]
\includegraphics[width=7cm,clip=true]{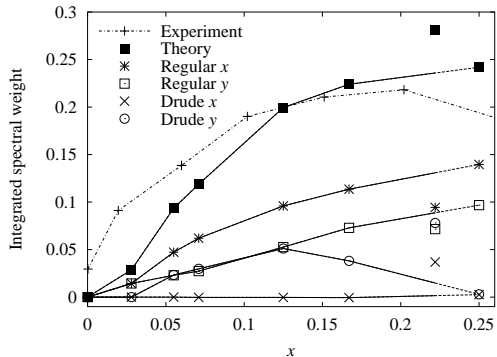}
\caption{Experimental\protect\cite{uch91} and
theoretical optical conductivity spectral weight
integrated up to 1.5 eV vs. $x$. The spectral weight is
converted to an effective number of electrons $N_{\rm eff}$
as in Ref.~\protect\cite{uch91}. The background dielectric constant
$\epsilon_b$ is our only free parameter and
was adjusted to make the theoretical and experimental intensities to
have an overall agreement.
 We also show the computed Drude and regular contributions in each 
direction. The dashed lines indicate the region of metastability
of the $d=4$ solution. The points not joined by lines correspond to
the $d=3$ solution which becomes more stable in that region.}
\label{fig:swdx}
\end{figure}

In Fig.~\ref{fig:swdx} we show the optical conductivity spectral
weight integrated up to an energy of 1.5 eV [$N_{\rm eff}$(1.5eV)]
in comparison with experimental data from Ref.~\cite{uch91}. The
lower curves report the regular ($\omega>0$) and Drude
($\omega=0$) contributions to $N_{\rm eff}$(1.5eV) in each
direction. 

For $x<1/8$ the number of stripes, which act as low-energy light 
absorbers,  increases linearly with doping keeping its 
electronic structure practically unchanged\cite{lor02}. 
Accordingly $N_{\rm eff}$(1.5eV) increases roughly 
linearly with doping.  For $1/8>x>x_1$ ($x_1\sim 0.22-0.23$)
the number of absorbers (stripes) get locked and the evolution 
of the spectral weight is related to changes in the electronic 
structure of each stripe. Especially the shift of the chemical potential from
the center of the active band\cite{lor02} leads to a depression of 
Drude weight in the stripe direction (joined circles)
and results in a slower increase of spectral weight with doping
which correlates well with experiment.

For $x>x_1$ $d=3$ stripes become the mean-field ground state. 
These solutions tend to have a larger spectral weight due to the
strong Drude weight in both directions 
(unjoined points in Fig.~\ref{fig:swdx}). 
The nature of the ground state however is not clear in this overdoped
regime and therefore our results become more qualitative than
quantitative. Indeed $d=3$ stripes have not been observed in 
LSCO and theoretically a combination of different solutions would 
likely be more appropriate\cite{lor02} which is beyond the scope of this work. 

To conclude we have computed the optical conductivity of metallic
stripes within a GA+RPA approach. 
Striped domain walls induce two MIR excitations: At around
1.3eV a HEMIR band appears which is related to interband
transitions within the stripe band structure. Further on we found
a collective mode (i.e. the LEMIR band) which softens as a
function of doping due to the suppression of the energy barrier between
quasidegenerate BC and SC stripe solutions. These features are in good
agreement with experiments for positions and relative intensities 
with parameters fixed by first principle computations as in 
Ref.~\cite{lor02}.     
It is worth speculating that scattering of holes moving along and
perpendicular to the stripes with the soft collective mode may be
responsible for both the anomalous normal state behavior and the
superconducting pairing in cuprates.

After this work was completed Ref.~\cite{hom02} was posted
in which by studding temperature effects, MIR features in isostructural 
 La$_{2}$NiO$_{4.133}$ were also linked to stripes. 

J.L. acknowledges hospitality at ICTP during this work and G.S.
acknowledges support from the Deutsche Forschungsgemeinschaft.

%\bibliographystyle{prsty}
%\bibliography{../../../tex/htsct,../../../tex/htsce,sigma}

\end{document}